\documentclass[twoside]{article}
\usepackage{graphicx}
\usepackage{fancyhdr}
\usepackage{multicol}
\usepackage{styl-ap}

\begin{document}
\title{Ultrahigh Energy Cosmic Rays: Review of the Current Situation}
\author{Todor Stanev\work{1}}
\workplace{Bartol Research Institute and Department of Physics 
 and Astronomy, University of Delaware, Newark, DE 19716, U.S.A}
\mainauthor{stanev@bartol.udel.edu}
\maketitle

\begin{abstract}
  We describe the current situation of the data on the highest energy
 particles in the Universe - the ultrahigh energy cosmic rays. The
 new results in the field come from the Telescope Array experiment
 in Utah, U.S.A. For this reason we concentrate on the results from
 this experiments and compare them to the measurements of the other
 two recent experiments, the High Resolution Fly's Eye and the 
 Southern Auger Observatory.
\end{abstract}


\begin{multicols}{2}
\section{Introduction}
 Two years ago I was asked to review at this meeting the new
 results of the measurements of the ultrahigh energy cosmic rays
 (UHECR). At that time there were two experiments that did
 such measurements: the High Resolution Fly's Eye (HiRes) in
 Utah, U.S.A., and the Auger Southern Observatory (Auger) in Mendoza,
 Argentina. HiRes is a detector that measures the fluorescent
 light emitted by the Nitrogen in the atmosphere when its atoms
 are excited by the numerous electrons of such large air showers.
 Its two fluorescent telescopes are able to detect showers that
 hit the ground up to distances of 40 km from the detectors.
 The two telescopes of HiRes can observe the air showers separately
 or in stereo mode with both telescopes.
 Auger is a hybrid experiment that combines four fluorescent 
 detectors (FD) with a huge surface array (SD) that covers 3,000 km$^2$.
 The surface array consists of 1,600 water Cherenkov tanks on a 
 triangular matrix with an average distance between the tanks
 of 1,500 m. The Cherenkov tanks are deep enough (almost three
 radiation lengths) to detect electrons, gamma rays, and muons,
 and thus measure the energy flow of the air shower. 

 A brief summary of the results at that time is that both detectors
 observed the GZK feature in the UHECR energy spectrum (Greisen 1966;
 Zatsepin\&Kuzmin 1966): the steep decline in the UHECR energy
 spectrum above energy of 4$\times$10$^{19}$ eV due to the
 energy loss in cosmic ray propagation from their presumably
 extragalactic sources to us. The two measured
 spectra have very similar shapes and agree with each other 
 within the systematic errors of about 20\%. The two experiments,
 however, did disagree on the chemical composition of UHECR:
 HiRes interpretation of the measured depth of shower maximum
 (X$_{max}$) and its fluctuations was that all UHECR are Hydrogen
 nuclei (protons) (Sokolsky, 2011), while
 Auger interpreted its results as a chemical composition becoming
 increasingly heavier with energy above 2$\times$10$^{18}$ eV
 (Kampert\&Unger, 2012). The interpretation of the 
 chemical composition from the $X_{max}$ measurement depends
 on the hadronic interaction model used which creates a significant
 systematic error. 

  Auger also saw a correlation of their
 highest energy events (above 55 EeV = 5.5$\times$10$^{19}$ eV)
 with nearby AGN and the smaller HiRes statistics did not show 
 any correlation. These results have not changed during the last
 two years.

\subsection{Telescope Array}
 The new results come from a new detector, the Telescope Array (TA),
 which is a hybrid detector that started collecting
 data in 2009 in Utah, USA, at 39$^o$N, 120$^o$W and altitude of
 1500 m a.s.l. Its surface array (SD) consists of 607 
 scintillator counters on a square grid with dimension of 1.2 km.
 Each scintillator detector consists of two layers of thickness
 1.2 cm and area of 3 m$^2$. The phototube of each layer is connected
 to the scintillator via 96 wavelength shifting fibers which make the
 response of the scintillator more uniform. Each station is
 powered by a solar panel that charges a lead-acid battery.
 The total area of the surface array is 762 km$^2$. The surface array is
 divided in three parts that communicate with three control towers 
 where the waveforms are digitized and triggers are produced.
 Each second the tower collects the recorded signals from all
 stations and a trigger is produced when three adjacent stations
 coincide within 8 $\mu$sec.The SD reaches a full efficiency 
 at 10$^{18.7}$ eV for showers with zenith angle less than
 45$^o$ (Nonaka 2009). This angle corresponds to SD 
 acceptance of 1,600 km$^2$sr.

  The fluorescence detector (FD) consists of three fluorescence stations.
 Two of them are new and consists of
 12 telescopes with field of view from elevations of 3$^o$ to 31$^o$.
 The total horizontal field of view of each station is 108$^o$.
 The third station has 14 telescopes that use
 cameras and electronics from HiRes-I and
 mirrors from HiRes-II. The fluorescent telescopes are calibrated 
 with N$_2$ lasers, Xe flashers, and an electron linear
 accelerator (Tokuno 2009).

 The atmosphere is monitored for clouds by IR cameras and with the
 use of the central laser facility which is in the center of the
 array at 20.85 km from each station. The fluorescent stations 
 are positioned in such a way that they cover the whole area of
 the surface detector. The mono acceptance of the FD is 1,830 km$^2$sr
 and the stereo one is 1040 km$^2$sr. The total energy resolution is 25\%
 and the $X_{max}$ resolution is 17 g/cm$^2$.

\begin{myfigure}
\centerline{\resizebox{80mm}{!}{\includegraphics{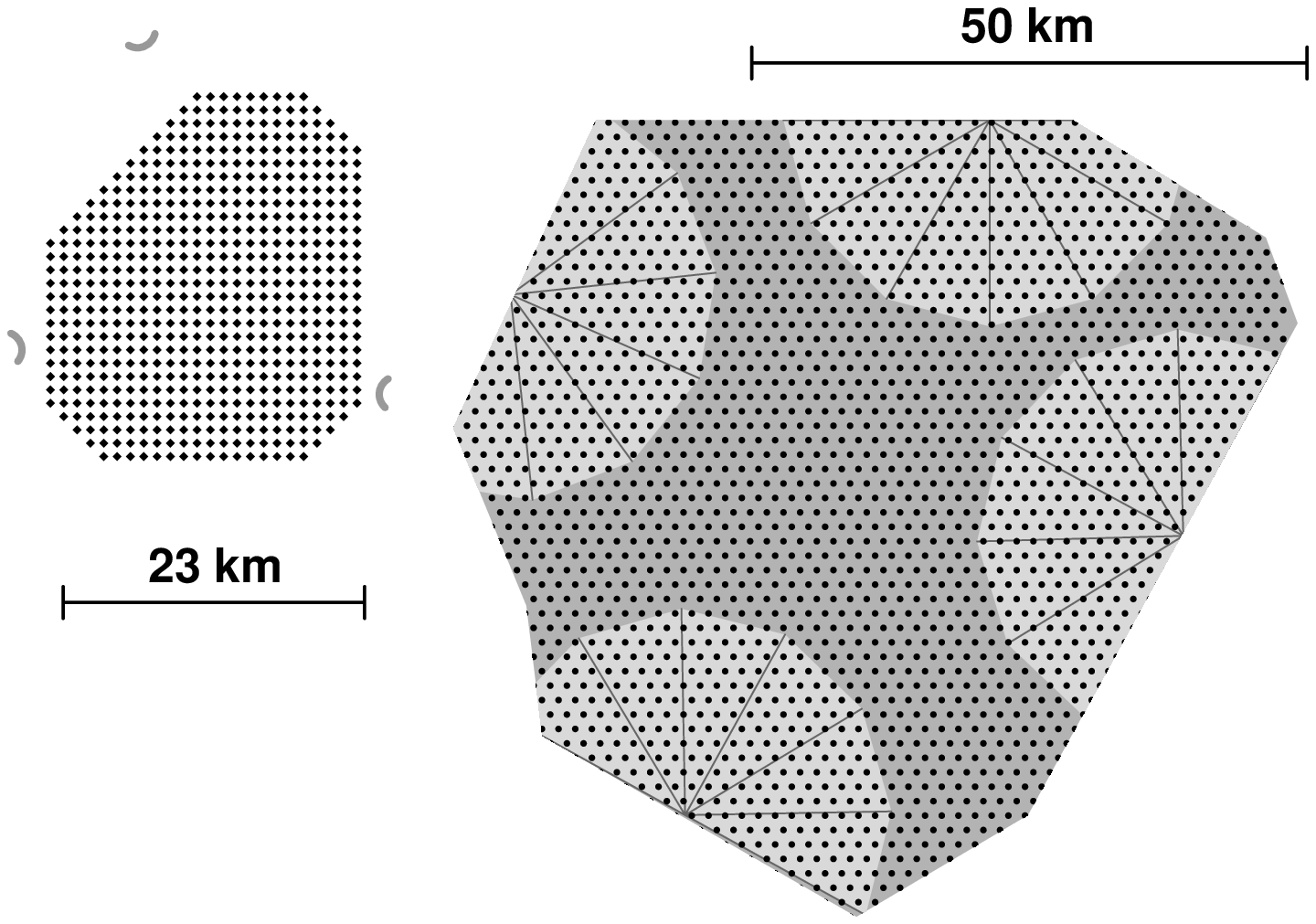}}}
\caption{Comparison of the sizes of the surface arrays of the Telescope
 Array and the Auger Southern Observatory. The position of the TA 
 fluorescent detectors are indicated with small arcs.}
\label{Auger_TA}
\end{myfigure}

\section{New results}
 The new results come from the Telescope Array. They were reported at
 the 2011 International cosmic ray conference in Beijing. Two papers
 also appeared in the arXiv a couple of months ago. Figure~\ref{Auger_TA}
 compares the size of the TA to that of Auger - it is almost four times
 smaller. In addition, the water Cherenkov tanks have the same effective
 area up to shower zenith angle of 60$^o$ which means that their 
 exposure is higher than that of the scintillator counters. For these 
 reasons the new TA results are based on smaller statistics and should
 be considered preliminary.

\subsection{UHECR energy spectrum}
 Figure~\ref{spee312} shows the energy spectrum measured by the 
 Telescope Array (Abu-Zayyad 2012a) compared to those of Auger
 and the HiRes experiments. At first glance at the figure we see
 that the spectrum measured by TA is extremely close to that of
 HiRes. One should say here that there is a big difference 
 between the way the energy spectrum is measured by the two detectors.
 The Telescope Array has used the method of measuring the energy spectrum
 with the surface array introduced by Auger. Fluorescent telescopes
 can work only in clear moonless nights with good atmospheric conditions
 (about 10\% of the time) while the surface arrays are active all
 the time. In addition, the energy estimates with the surface array
 depend heavily on the hadronic interaction model used in the
 shower analysis. To increase the statistics one can correlate
 the particle density in the surface array at certain distance
 from the shower core (800 m for TA and 1,000 m for Auger) with
 the energy estimate from the fluorescent detectors (which does not
 need the hadronic Monte Carlo) and then use the surface density 
 to obtain the spectrum. 

\begin{myfigure}
\centerline{\resizebox{80mm}{!}{\includegraphics{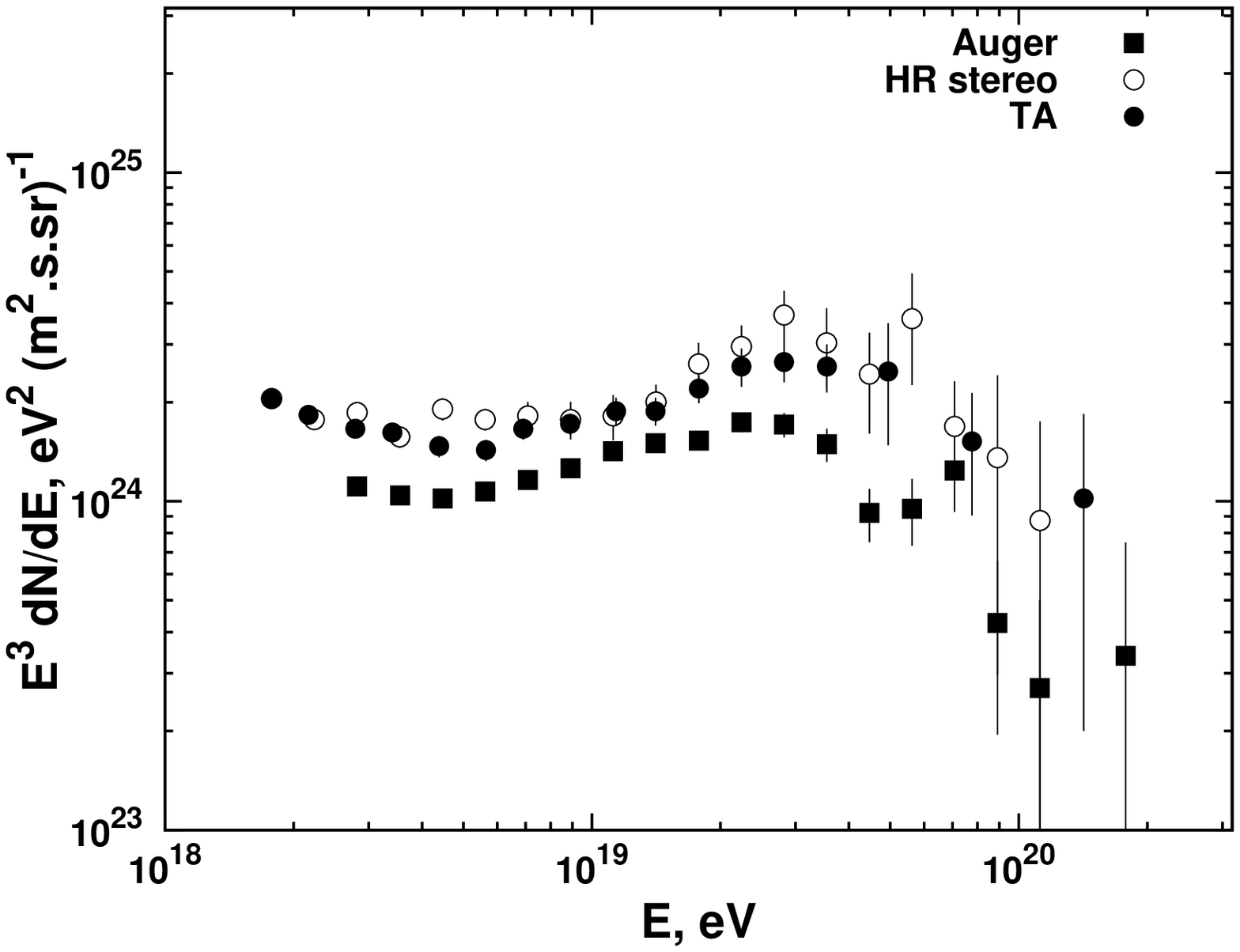}}}
\caption{Energy spectrum of the UHECR measured by TA, HiRes and Auger.
 The particle flux is multiplied by E$^3$ to show better the shape of
 the energy spectrum.}
\label{spee312}
\end{myfigure}
 
 The Telescope Array energy spectrum paper (Abu-Zayyad 2012a) also 
 fits the shape of the spectrum with a broken power law. 
 The {\em ankle} of the spectrum, where it becomes less steep, is
 at (4.8$\pm$0.1)$\times$10$^{18}$ eV. The power law index $\alpha$
 before the ankle is 3.33$\pm$0.04, at the ankle it is 2.68$\pm$0.04
 and at the GZK decline it is 4.2$\pm$0.7. The statistics is, of 
 course, quite small but there is no doubt that the spectrum becomes
 steeper as predicted by Greisen and Zatsepin\&Kuzmin.
 It is indeed remarkable that using very different methods for
 observation of the spectrum the data of TA and HiRes agree so well. 

 One has to admit that the shape of the energy spectrum detected
 by TA is also very similar to that of Auger in spite of the 
 different normalization. All three spectra shown
 if Fig.~\ref{spee312} are consistent within the systematic
 errors claimed by the experiments which are of order 20\%.

\subsection{Chemical composition of UHECR}
  The measurement of the chemical composition of cosmic rays is 
 through the interpretation of the depth of shower maximum $X_{max}$.
 The position on the shower maximum for proton showers becomes
 deeper in the atmosphere with energy because showers continue
 developing until the average energy of its particles decreases
 below 80 MeV.
 Showers caused by heavy nuclei have $X_{max}$ higher in the atmosphere
 because in the first approximation they are the sum of A nucleon 
 showers of energy E/A. At energies above 10$^{18}$ eV the difference
 between $X_{max}$ of proton and iron showers is about 100 g/cm$^2$. 
 The primary mass of the particle interacting in the atmosphere 
 also affects the fluctuations of $X_{max}$ per energy bin. Showers
 caused by heavy nuclei would have smaller fluctuations as in the
 simplest model (superposition) the fluctuations in such showers
 should decrease by $\sqrt(A)$. In Monte Carlo calculations the
 difference is smaller varying from about 60 g/cm$^2$ for proton
 showers to about 20 g/cm$^2$ for Fe showers.

 Figure~\ref{compos} compares the $X_{max}$ measurements of the Telescope
 Array (Tsunesada 2011) presented in the 2011 International Cosmic Ray
 Conference (Beijing) to the results of HiRes and Auger.
\begin{myfigure}
\centerline{\resizebox{80mm}{!}{\includegraphics{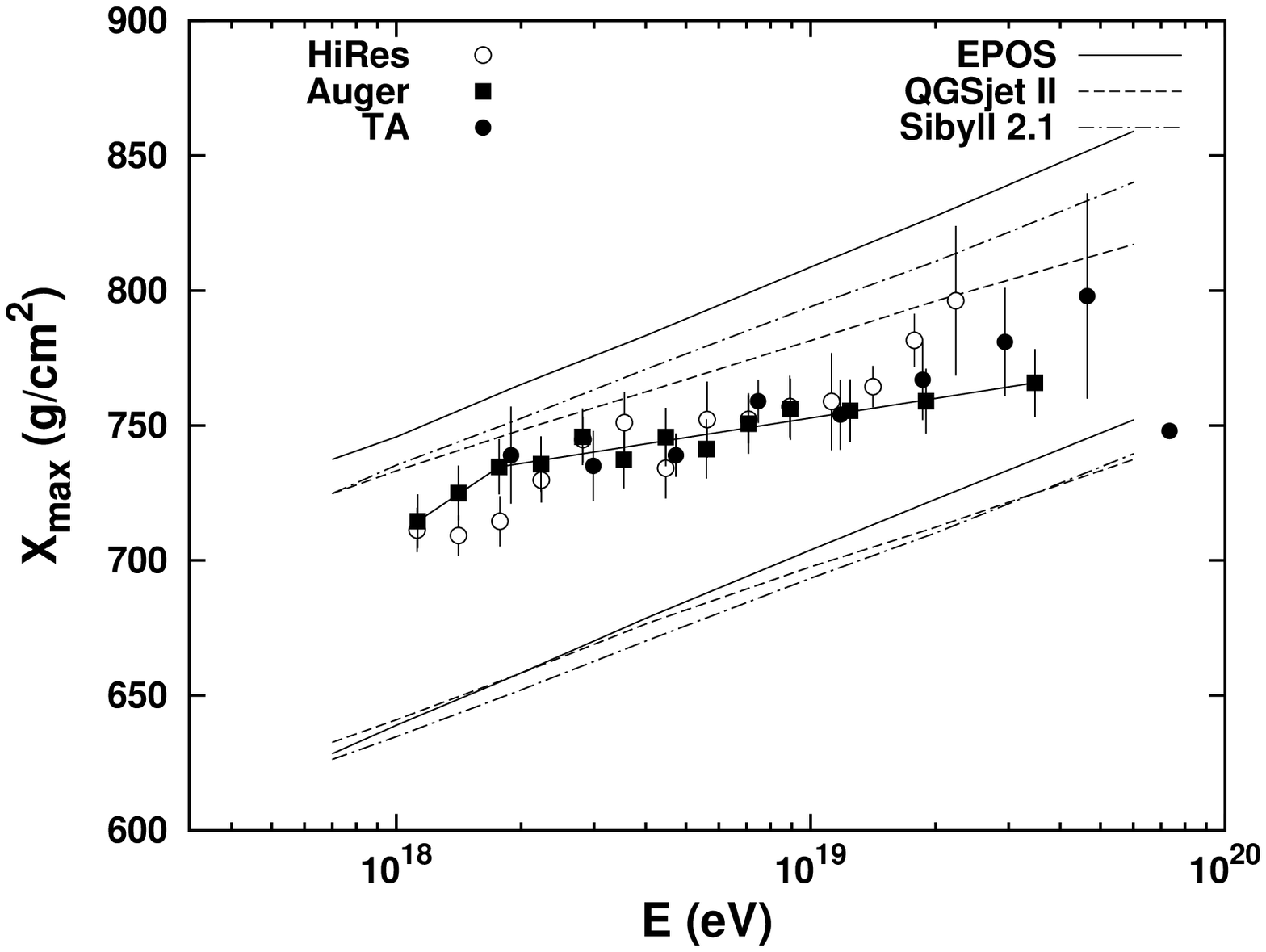}}}
\caption{Depth of shower maximum measurements by the Telescope 
Array, HiRes and Auger. The lines show the energy behavior for proton
and iron showers for three hadronic interaction models.}
\label{compos}
\end{myfigure}

 The interpretation of the $X_{max}$ measurement by the TA experiment
 is that the UHECR composition is light, consisting mostly of 
 protons and very light nuclei. It is not easy to understand the
 very different interpretations of the HiRes and TA (on one hand) and
 Auger of the data, which look very similar to the naked eye. 
 The explanation of the previous disagreement between HiRes and Auger
 was that they used different event selection. 
 It is not obvious now what exactly is the TA event selection. One has 
 to have in mind that the highest energy two points in its data set
 have respectively only three and one events and the average $X_{max}$
 could be different when more statistics is collected.

 The Telescope Array also presented (Tsunesada 2011) the distributions
 of $X_{max}$ in the energy bins shown in Fig.~\ref{compos}. 
 At relatively low energy the width of the distributions were more
 similar to proton showers, while at high energy the statistics is
 not enough to judge the distributions. 

\subsection{Identifying the sources of UHECR}

 In 2007 the Auger Collaboration published a paper where a correlation
 of their highest energy events ($\rangle$ 55 EeV) with AGN was discussed.
 At that time the collaboration has seen 27 such events. Eighteen of
 these events had an angle of less than 3.2$^o$ from the positions
 of nearby (redshift $z$ < 0.018, distance less than 75 Mpc) AGN from
 the V\'{e}ron-Cetty and V\'{e}ron catalog (VCV) (V\'{e}ron-Cetty 2006).
 The correlation was even stronger if events 
 close to the galactic plane were excluded. Although the VCV catalog
 contains mostly not very powerful Seyfert-2 AGN they may have marked
 the the distribution of the real sources. This paper had a huge
 readership and many scientists were convinced that the sources of
 UHECR would be discovered soon. The HiRes data (13 events) did not
 confirm  this correlation (Abbasi 2008) and papers discussing the
 different fields of view (Auger in the South and HiRes in the North)
 appeared in press.

 Since at that time the Southern Auger Observatory was completed it
 did not take a long time to significantly increase the statistics.
 In 2009 the correlation of 69 high energy events with the same 
 AGN catalog was published. The correlation has decreased to about
 38\% of the events. The previous result happened to be a typical
 3$\sigma$ disappearing result. 

 The disagreement between Auger and HiRes on the correlation of the
 arrival directions of their highest energy events with AGN is also
 strange because of their results on the chemical composition of UHECR.
 If the composition is indeed  heavy, as interpreted by Auger, one
 expects that the heavy nuclei would scatter more in the intergalactic
 and galactic magnetic fields and show no anisotropy.

\begin{figure}[htb]
\centerline{\resizebox{150mm}{!}{\includegraphics{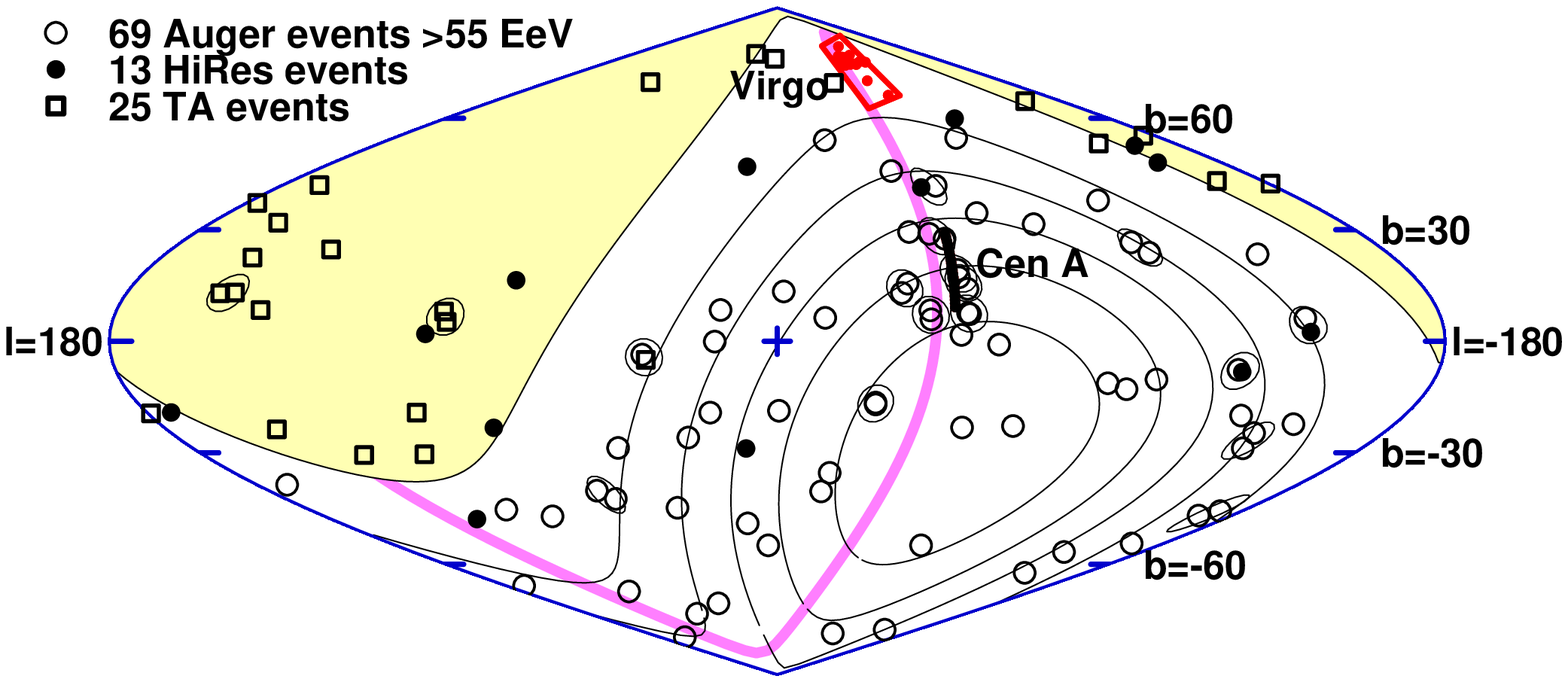}}}
\caption{Arrival directions of the 69 Auger events, 13 HiRes
events and the TA 25 events in galactic coordinates. The colored area
shows the part of the Galaxy that Auger does not see. The six areas defined
within the Auger field of view have equal exposures. The events that form a
pair at angular distance less than 5$^o$ are circled.}
\label{corr}
\end{figure}

 Figure~\ref{corr} shows the arrival directions of the highest energy
 events of Auger, HiRes and TA. Having in mind the dimensions of 
 Auger and TA (see Fig.~\ref{Auger_TA}) and the fact that TA field
 of view is restricted to zenith angles less than 45$^o$ it
 is difficult to believe
 that the ratio of their statistics is less than three. We hope that
 Auger has more than 100 such events by now. The 20\% difference in
 the energy assignment may also play a role in this issue.

\end{multicols}

\begin{multicols}{2}

 It is not easy to judge what the new data set says about the correlation
 of the UHECR arrival direction with powerful astrophysical sources.
 One way would be to judge the possible direction of the sources by
 close-by arrival directions of groups of highest energy events.
 We looked at pairs of events at angular distance less than 5$^o$ from
 each other. There are 11 such pairs in the Auger 69 events 
 data set. Six such pairs are within 18 degrees of CenA. An isotropic 
 Monte Carlo in the Auger field of view creates on the average 11 pairs,
 the same number as in data. There are three pairs consisting on HiRes
 and Auger events and one TA-Auger pair. There also two pairs consisting
 of TA events as shown in Fig.~\ref{corr}. It is not possible to run 
 an isotropic Monte Carlo for the new events because the exposure of
 the Telescope Array is not as well defined as those of Auger and HiRes.
 
\section{Discussion}
  It is not possible to conclude anything new from the data set of the
 Telescope Array. Its results on the energy spectrum of the UHECR is
 very similar to that of the High Resolution Fly's Eye. All three 
 newest experiments confirm the end of the cosmic ray spectrum that
 is consistent with the GZK effect and with photo dissociation energy
 loss of heavy nuclei. The three published spectra are
 almost identical within the stated systematic error of more than 20\%.
 It may be important for high energy physics to understand the differences,
 claimed by Auger and TA, between the energy assignment of the events
 from the fluorescent detectors and the surface arrays, which is also of order
 20\%. The analysis of the surface array data in both detectors give
 a higher energy assignment.

 In the case of Auger the suspicion is that the water Cherenkov tanks
 of the surface array see a much higher number of muons, which produce
 more light in the tanks than electrons and $\gamma$-rays do.
 In the case of TA the surface array consists
 of scintillator counters where muons generate the same signals as
 electrons do. In this case a wrong expectation about the shower muons
 would have smaller contribution to the energy assignment.  

 By far the biggest controversy in the results is the interpretation
 of the $X_{max}$ measurement by the three experiments shown in
 Fig.~\ref{compos}. The results of the measurements do not seem to 
 be as different to the eye as the interpretation is. HiRes and TA
 interpret the results as almost purely proton composition while 
 Auger interprets the measurements as a composition becoming 
 increasingly heavier with energy. In the review of UHECR
 (Letessier-Selvon 2011) the suspicion was on the different event
 selection in Auger and HiRes. We do not know much about the
 selection in TA yet and this question is still open. 

 There is some theoretical contradiction between the chemical composition
 derived by Auger and the anisotropy it has measured, including the large
 number of events coming from the vicinity of CenA. Lemoine \& Waxman
 (Lemoine 2009) 
 suggested that if the composition were heavy there would be protons from
 nuclear photodissociation that would show the same anisotropy at 
 significantly lower energy. Such anisotropy at about 10$^{18}$ eV has
 not been seen by the Auger experiment. This is not an argument against
 the heavy composition derived by Auger, but an interesting argument 
 for further measurements and observations.

 The new data on the arrival direction distribution of UHECR that come
 from TA did not contribute to the source identification. It is very good
 though, to have an active experiment in the Northern Hemisphere.
 Auger and TA are able to increase the statistics by a factor close to
 five during the next four years. This statistics may not be sufficient
 for the identification of the sources of the ultrahigh energy cosmic
 rays, but will certainly be an improvement over the current 
 situation. 

 The good news is that at the International Symposium on Future 
 Directions in UHECR physics at CERN in February 2012 the two
 collaborations have started to work together on all of the topics
 discussed above. Working groups consisting of members of both
 collaborations were created and gave talks at the symposium. 
 All of us hope that the working groups will study well the 
 differences in the shower reconstruction and data analysis and
 will at least discover the reasons for the contradictory results.
 If this happens we will know much more about this exciting field 
 in a couple of years. 

\thanks The author thanks the organizers of the Vulcano workshop 
 for the invitation to this excellent and useful meeting. His work 
 is supported in part by the DOE grant DE-FG02-91ER40626.

\bigskip
\bigskip
\noindent {\bf DISCUSSION}

\bigskip
\noindent {\bf PETER GRIEDER:} Concerning the differences in the 
 composition between Auger and the Telescope Array. The two experiments
 see different sources. These maybe of different nature. Please comment.

\bigskip
\noindent {\bf TODOR STANEV:} The fields of view of Auger and TA are
 different but it is difficult to imagine that the cosmic ray composition
 that much. The fields of view of Auger and HiRes coincided about 30\%
 so HiRes should have seen some heavy nuclei. I do not believe that this
 the reason for the disagreement.

\bigskip
\noindent {\bf LAURENCE JONES:} We now know that the total p-p cross section
 rises to about 100 mb near 1 EeV. Do the Monte Carlo models used to 
 determine the mass include the cross section rise?

\bigskip
\noindent {\bf TODOR STANEV:} The hadronic Monte Carlo models used for
 shower analysis have rising cross section. The cross section of SIBYLL 2.1
 is higher than the one measured at LHC. All interaction models are now 
 revised to match the measurements.

\bigskip
\noindent {\bf ANATOLY ERLYKIN:} Will the extreme sharpness of the ankle
 in the published Telescope Array surface array energy spectrum is 
 evidence against the dip model of its origin?

\bigskip
\noindent {\bf TODOR STANEV:} The first point of the TA energy spectrum is
 indeed quite high. Since it is only one point at the detector threshold,
 where the detector is not fully efficient, I have not paid much attention
 to it.
 
\end{multicols}

\begin{thebibliography}{99}
\bibitem{} Abbasi, R.U. et al (HiRes Collaboration), 2008, Astropart. Phys.
 {\bf 30}, 175.

\bibitem{} Abraham, J. et al (Auger Collaboration), 2007, Science, {\bf 318}
 938

\bibitem{} Abu-Zayyad, T. et al (Telescope Array), arXiv: 1205.5067

\bibitem{} Abu-Zayyad, T. et al (Telescope Array), arXiv: 1205.5984

\bibitem{} Greisen, K., Phys. Rev. Lett., 1966 {\bf 16} 748

\bibitem{} Kampert, K-H. and Unger, M., Astropart. Phys. 2012 {\bf 35} 660

\bibitem{} Lemoine, M. \& Waxman, E., JCAP 2009 {\bf 0911} 009

\bibitem{} Letessier-Selvon, A. \& Stanev,T., Rev. Mod. Phys. 2011 {\bf 83} 907

\bibitem{} Nonaka, T. et al (Telescope Array), Nucl. Phys. B (Proc. Suppl.),
 2009, {\bf 190} 26

\bibitem{} Sokolsky, P. (HiRes Collaboration), Nucl. Phys. B (Proc. Suppl),
 2011, {\bf 212-213} 74

\bibitem{} Tokuno, H. et al (Telescope Array), Nucl. Instrum. Meth.,
 2009 {\bf A601} 364

\bibitem{} Tsunesada, Y.: in Proceedings of the 32nd ICRC, Beijing, 2011,
 {\bf 12} 58

\bibitem{} V\'{e}ron-Cetty, M.-P. and V\'{e}ron, P., 2006,
 Astron\&Asttrophys, {\bf 445} 773

\bibitem{} Zatsepin, G.T. and Kuzmin, V.A., JETP Lett., 1966 {\bf 4} 78 

\end{thebibliography}
\end{document}